\documentclass{aa}  
\usepackage{graphicx}
\usepackage{txfonts}
\begin{document}
   \title{Mass loss out of close binaries}
   
   \subtitle{The formation of Algol-type systems, completed with case B RLOF}

   \author{W. Van Rensbergen, J.P. De Greve, N. Mennekens, K. Jansen \and C. De Loore}
 
  \institute{Astrophysical Institute, Vrije Universiteit Brussel, Pleinlaan 2, 1050 Brussels, Belgium\\
   \email {wvanrens@vub.ac.be}
   }

    \date{Received August 16, 2010; accepted December 27, 2010}

 
  \abstract
   {Liberal evolution of interacting binaries has been proposed previously by several authors in order to meet various observed binary characteristics better than conservative evolution does. Since Algols are eclipsing binaries the distribution of their orbital periods is precisely known. The distribution of their mass ratios contains however more uncertainties. We try to reproduce these two distributions theoretically using a liberal scenario in which the gainer star can lose mass into interstellar space as a consequence of its rapid rotation and the energy of a hot spot.}
   {In a recent paper (Van Rensbergen et al.~2010,~A$\&$A) we calculated the liberal evolution of binaries with a B-type primary at birth where mass transfer starts during core hydrogen burning of the donor. In this paper we include the cases where mass transfer starts during hydrogen shell burning and it is our aim to reproduce the observed distributions of the system parameters of Algol-type semi-detached systems.}
   {Our calculations reveal the amount of time that an Algol binary lives with a well defined value of mass ratio and orbital period. We use these data to simulate the distribution of mass ratios and orbital periods of Algols.}
   {Binaries with a late B-type initial primary hardly lose any mass whereas those with an early B primary evolve in a non-conservative way. Conservative binary evolution predicts only $\sim$ 12 $\%$ of Algols with a mass ratio $q$ above 0.4. This value is raised up to $\sim$ 17 $\%$ using our scenario of liberal evolution, which is still far below the $\sim$ 45 $\%$ that is observed.}
   {Observed orbital periods of Algol binaries larger than one day are faithfully reproduced by our liberal scenario. Mass ratios are reproduced better than with conservative evolution, but the resemblance is still poor.}

   
    \authorrunning{W. Van Rensbergen et al.}  
     \titlerunning{Mass loss out of close binaries}
   \maketitle

\section{Introduction}

Warner (\cite{Warner}) introduced the denomination $liberal$ to distinguish binary evolution with a mass loss and subsequent angular momentum loss from the $conservative$ case where no mass leaves the system. Van Rensbergen et al. (\cite{Walter2}) developed a liberal scenario in which mass can be lost from a binary during a short era of rapid mass transfer soon after the onset of Roche lobe overflow (RLOF). The joint catalog with evolutionary tracks was extended with the evolutionary tracks for binaries starting from sufficiently small orbital periods, so that RLOF starts during core hydrogen burning of the donor (case A) as presented by Van Rensbergen et al. (\cite{Walter3}). In this contribution these calculations have been extended towards larger initial orbital periods so that RLOF starts after exhaustion of hydrogen in the core of the donor (case B). Our catalog (Van Rensbergen et al. \cite{Walter2}) contains now 561 conservative and liberal evolutionary tracks and is available at the Centre de Donn\'ees Stellaires (CDS). The grid of calculations covers binaries with a B-type primary at birth.

A binary is considered to be an Algol when the semi-detached system shows the typical characteristics as mentioned by Peters (\cite{Peters}), where the less massive donor fills its Roche lobe, the more massive gainer does not fill its Roche lobe and is still on the main sequence and the donor is the cooler, fainter and larger star. The distribution of orbital periods of Algols is precisely known since the time between two subsequent eclipses for every system has been measured accurately. The catalog of Budding et al. (\cite{Budding et al.}) extended with the semi-detached Algols from Brancewicz et al. (\cite{Brancewicz et al.}) supplies us with 303 Algols which can be issued from binary evolution with a B-type primary at birth. Van Rensbergen et al. (\cite{Walter1}) showed that the distribution of mass ratios strongly depends on the method used to interpret the observations. Substantial differences are found between mass ratios derived so as to make the parameters of the most massive star fit main sequence characteristics ($q_{MS}$) and mass ratios $q_{LC}$ and $q_{SD}$ which are respectively obtained by the light curve solution and using the fact that an Algol is a semi-detached system. In this paper we used the observed mass ratio distribution of Van Rensbergen et al. (\cite{Walter2}) who applied a mix of $q_{MS}$ and $q_{LC}$ which represents the numbers in the Pourbaix et al. catalog (\cite{Pourbaix}) best.

Mass loss during liberal evolution is defined by a quantity $\beta$ $\in$ [0,1] giving the fraction of mass lost by the donor (subscript $d$) that is accreted by the gainer (subscript $g$):

\begin{equation}
{\dot M_{g} = -~ \beta ~\dot M_{d}^{RLOF}}
\label{Beta} 
\end{equation}

\vspace{1mm}

Conservative evolution is characterized by $\beta$ = 1. Liberal theoretical calculations (i.e. those in which, at least for some time, $\beta < 1$) depend not only on the amount of mass lost from the system (related to the parameter $\beta$) but also on the amount of angular momentum taken away by this matter (related to the parameter $\alpha$), formulated by the  $\alpha$, $\beta$-mechanism of Podsiadlowski et al. (\cite{Podsiadlowski}).
The quantity $\alpha$ is then defined by Rappaport et al. (\cite{Rappaport}) so that, for the angular momentum loss rate from the system $\dot{J}$ in the case of mass loss rate of the donor $\dot{M}_{d}^{RLOF}$ one has that:

\begin{equation}
\dot{J}=\alpha~\dot{M}_{d}^{RLOF}~(1-\beta)~\frac{2~\pi~a^{2}}{P}
 \label{Jdot} 
\end{equation}

with $a$ and $P$ respectively the semi-major axis and orbital period of the binary system. The value of $\alpha$ thus depends on how much angular momentum is carried away from the system by a certain amount of mass loss, i.e. on the way in which matter leaves the binary.

Values of $\beta$ $<$ 1 were proposed by Meurs $\&$ Van den Heuvel (\cite{Meurs}) in order to theoretically reproduce the observed numbers of persistent strong massive binary X-ray sources and the number of observed Wolf-Rayet binaries. Binary evolutionary calculations with constant values of $\beta$ (e.g. $\beta$ = 0.5) were then reproduced by various authors, e.g. De Loore $\&$ De Greve (\cite{Degreve}) and more recently by De Mink et al. (\cite{Demink}) ($\beta$ = 0.25, 0.5 and 0.75).

\vspace{1mm}

Although mass loss from binaries is needed in evolutionary theory, it would be astonishing that $\beta$ does not depend on the mass transfer rate. It is more plausible that evolution of semi-detached binaries remains conservative ($\beta$ = 1) during the long lasting quiet eras of slow RLOF, and that mass can only be lost from the system ($\beta$ $<$ 1) during short lasting violent eras of rapid mass transfer.

The stellar radiation and the spin-up of the gainer as given by Packet (\cite{Packet}) is used by Langer (\cite{Langer}) to compute the mass loss rate for massive binaries. This scenario yields a time dependent behavior of $\beta$ as published by Wellstein et al. (\cite{Wellstein}) and Petrovic et al. (\cite{Petrovic}).

The stellar radiation of a rapidly rotating intermediate mass gainer is however not sufficient to drive mass out of the system. Hence we introduced in the $Brussels$ binary evolution code the possibility of mass loss from a binary as a consequence of spin-up and hot spots created on the gainer by mass transferred from the donor. The physics of this hot spot model, as well as the resulting mass loss from a binary with a B-type primary at birth were discussed by Van Rensbergen et al. (\cite{Walter2,Walter3}) using this scenario.
     

\section{Method}

\subsection{Brussels evolutionary code}

Our results were obtained using the $Brussels$ binary evolutionary code. A detailed description of this code can be found in De Loore $\&$ Doom (\cite{Deloore}). It is in fact a solution of the stellar structure equations, introduced by Henyey et al. (\cite{Henyey1, Henyey2}) and adapted by Kippenhahn $\&$ Weigert (\cite{Kippenhahn1}) and Kippenhahn et al. (\cite{Kippenhahn2}) for binary evolution. The calculations were performed with the OPAL-opacities (Rogers \& Iglesias \cite{Rogers}, Iglesias et al. \cite{Iglesias1}, Iglesias \& Rogers \cite{Iglesias2}).

For the input physics, convective mixing, radius corrections and nuclear physics, we refer to Prantzos et al. (\cite{Prantzos}). Moderate convective core overshooting is applied, as described in De Loore \& De Greve (\cite{Degreve}). Mass loss by stellar wind is included using the semi-empirical formalism of De Jager et al. (\cite{Dejager}), and the period changes due to subsequent angular momentum loss are also taken into account. However, these small losses are of little importance to the evolution outcome. For the spin-up of the gainer star and the evolution of the orbital separation, an angular momentum balance is considered, which includes the orbital angular momentum of the system and the spin angular momenteum of the gainer. The latter is calculated assuming solid-body rotation. When the gainer is spun up, its increase in spin angular momentum is extracted from the orbit. The spin angular momentum of the donor, which is small compared to the others, is currently not taken into account, but will be in future research.

\subsection{Initial conditions for binaries with a B-type primary}
\label{sec_Initial}

Van Rensbergen et al. (\cite{Walter1}) used non-evolved systems in the 9th catalog of Spectroscopic Binaries of Pourbaix et al. (\cite{Pourbaix}) to establish the initial conditions for the evolution of binaries with a B-type primary at birth. We distinguished between late B-type primaries in the mass-range [2.5,7] $M_{\odot}$ and early B-type primaries in the range [7,16.7] $M_{\odot}$. The following initial conditions were adopted:

\begin{flushleft}
$\bullet$~~future donors follow a normalized IMF: $\xi$($M_{d}$)=C$~$$M_{d}^{-x}$~~with $x$=2.35 as given by Salpeter  (\cite{Salpeter}).\\ 
$\bullet$~~binaries with a B-type primary at birth have a normalized bimodal period distribution $\Pi$(P)=$\frac{A}{P}$ (with $A$ such that the distribution is normalized over the entire period range) for their initial orbital periods (Van Rensbergen et al. \cite{Walter1}): 79.3~$\%$ of the systems with a late B primary follow a distribution as proposed by Popova et al. (\cite{Popova}) from $P_{min}$ = 0.93 d to $P_{max}$ = 33.37 d, whereas the remaining 20.7~$\%$ follow a similar distribution between $P_{min}$ = 33.37 d and $P_{max}$ = 9000 d. Out of the systems with an early B primary at birth, 79.1 $\%$ follow a Popova distribution  from $P_{min}$ = 1.02 d to $P_{max}$ = 12.91 d, whereas the remaining 20.9 $\%$ follow a similar distribution between $P_{min}$ = 12.91 d and $P_{max}$ = 4000 d. Such a bimodal distribution has also been found by Sana et al. (\cite{Sana}) for the initial distribution of orbital periods of massive binaries in NGC~6611, where they found 60 $\%$ between $P_{min}$ = 2 d and $P_{max}$ = 10 d, whereas the remaining 40 $\%$ have initial orbital periods between $P_{min}$ = 10 d and $P_{max}$ = 3200 d.\\
$\bullet$~~initial mass ratios ($q$=${M_{g}\over M_{d}}$) follow a normalized distribution: $\Psi$($q$)=C$~$$(1+q)^{-\delta}$~;~$\delta$=0.65 for non-evolved binaries with a late B primary and $\delta$=1.65 for non-evolved binaries with an early B primary, as given by Van Rensbergen et al. (\cite{Walter1}).\\  
\end{flushleft}

\subsection{Liberal evolution of binaries}

We completely followed the scenario as outlined in Van Rensbergen et al. (\cite{Walter3}). The details of the scenario can be found in that paper. During mass transfer the gainer is spun up and a hot spot is created in the accretion region which is located at the surface of the gainer in the case of direct impact or at the edge of the accretion disk if the criterion of Lubow $\&$ Shu (\cite{Lubow}) tells us that such a disk is created around the gainer.

\subsubsection{Spin-up of the gainer}

\begin{figure*}[!ht]
\centering
\includegraphics[width=9.6cm]{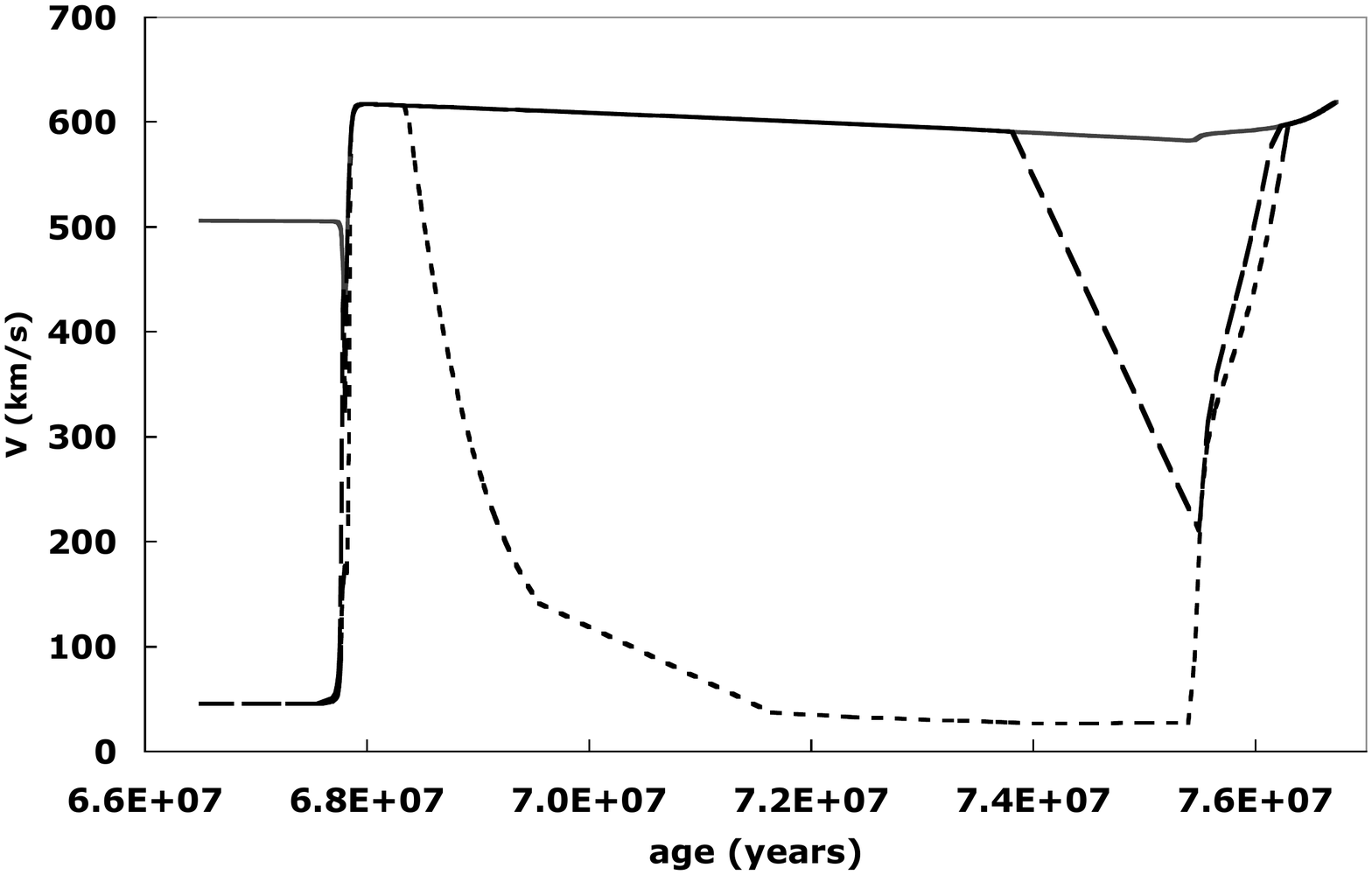}
\caption{Equatorial velocity of the gainer of a (6+3.6)$M_{\odot}$ binary with an initial orbital period of 3 days. Strong tidal interaction is represented by a dashed line, weak tides by a long-dashed one. The critical velocity is indicated by a solid line.} 
\label{fig_fig1}
\end{figure*}

Disregarding tidal interactions, Packet (\cite{Packet}) showed that the spin-up of the gainer goes very fast during the short era of rapid mass transfer soon after the onset of RLOF. Figure  {\ref{fig_fig1}} shows the spin-up for the gainer of a  (6+3.6) $M_{\odot}$ binary with an initial orbital period of 3 days. The spin-up was calculated with both strong and weak tidal interaction. The formalism was taken from Zahn (\cite{Zahn}). Strong tidal interaction is calculated with $f_{sync}$~=~0.1, whereas $f_{sync}$~=~1 implies weak tides, as adopted by Langer et al. (\cite{Langer2}) and discussed in detail by Wellstein (\cite{Wellstein2}). One sees in Fig. {\ref{fig_fig1}} that the gainer is spun up to critical velocity very quickly during the rapid era of mass transfer during core hydrogen burning of the donor. This is the case for slow tidal interaction as well as for a binary undergoing strong tides. During the slow phase of mass transfer the rotation of the gainer synchronizes rapidly in the case of strong tidal interaction. In the case of weak tidal interaction, synchronization is never achieved before the onset of RLOF B. The rotation of the gainer spins up again towards critical velocity when a second era of fast mass transfer occurs during hydrogen shell burning of the donor. It is clear that with rapid rotation a favorable situation is created for the gainer to lose mass into interstellar space. The example shown in Fig. {\ref{fig_fig1}} will turn out to evolve marginally $liberal$: no mass will be lost from the system during core hydrogen burning of the donor whereas only $\Delta$$M$ = 0.006 $M_{\odot}$ will be lost during hydrogen shell burning of the donor due to the combined action of rapid rotation and the creation of a hot spot on the gainer's surface.

\subsubsection{The hot spot}

The expression $L_{acc}^{\infty}$ = G$~ \frac {M_{g}~ {\dot{M}_{d}^{RLOF}}}{R_{g}}$ is often used to calculate the accretion luminosity. This expression is only valid in the case of matter coming from infinity impacting on the gainer's surface. In the case of interacting binaries the transferred matter is however issued from the first Lagrangian point $L_{1}$ and not from infinity. Hence, the real accretion luminosity has to be multiplied by a factor $D$ ($D \in [0,1]$): $L_{acc}~=~D ~ L_{acc}^{\infty}$. The geometrical factor $D$ will be large when $L_{1}$ is far away from the point of impact $P$ but can also be very small for binaries which are almost in contact. The factor $D$ is the geometrical factor included in the last column of Table {\ref{tab_tab1}}. The factor $S$ is the fractional surface area of the hot spot. This factor has been observed in some cases and Pringle (\cite{Pringle}) derived a theoretical expression\footnote{containing as a printing error, which has to be corrected when applying this formula, an exponent (-0.5) instead of (0.5) for $T_{eff}$.} starting from the characteristics (surface and opening angle) of the opening in the first Lagrangian point $L_{1}$ and the fact that the stream keeps its shape until it arrives at the point of impact $P$:

\begin{equation}
 S= 0.4556 ~10^{-4} ~ T_{eff,d}^{0.5}~ \frac{\left\vert L_{1}P\right\vert}{R_{\odot}}  ~
\left ( \frac{R_{d}}{R_{\odot}} \right)^{\frac{3}{2}}~\left (\frac{M_{d}}{M_{\odot}} \right)^{-\frac{1}{2}}~
\left (\frac{R_{g}}{R_{\odot}}\right)^{-2}.
 \label{SPringle} 
\end{equation}

\vspace{0.1cm}

The efficiency of the liberal scenario is further defined by the value of the factor of radiative efficiency $\tilde{K}$ which defines the fraction of $L_{acc}$ that is used by the hot spot as radiation pressure. This accretion luminosity is weakened on the one hand by the fact that only a fraction can be converted into radiation (represented by a weakening factor $A > 1$: $L_{add}$ = $\frac{L_{acc}}{A}$) and on the other hand strengthened because the energy of the impacting material is concentrated in a hot spot which is significantly smaller than the entire gainer's surface (represented by the fractional surface ($S=\frac {S_{spot}}{S_{g}}<1$). The radiation pressure of the hot spot is thus supplied by

\begin{equation}
\frac{L_{acc}}{A~ S}=L_{acc} ~  \tilde{K}.
 \label{Lacc} 
\end{equation}

Van Rensbergen et al. (\cite{Walter3}) showed that in the case of direct impact on the gainer's equator, the quantity $\tilde{K}$ can be calculated as follows:

\begin{equation}
 \tilde{K}~=~{{{\left (\frac {R_{g}} {R_{\odot}} \right )}^2}~ {\left (T_{spot}^{4}-T_{eff,g}^{4} \right )} \over \frac{L_{acc}}{L_{\odot}}~ (5770)^{4}}.
 \label{Ktilde} 
\end{equation}

\begin{figure*}[!ht]
\centering
\includegraphics[width=9.6cm]{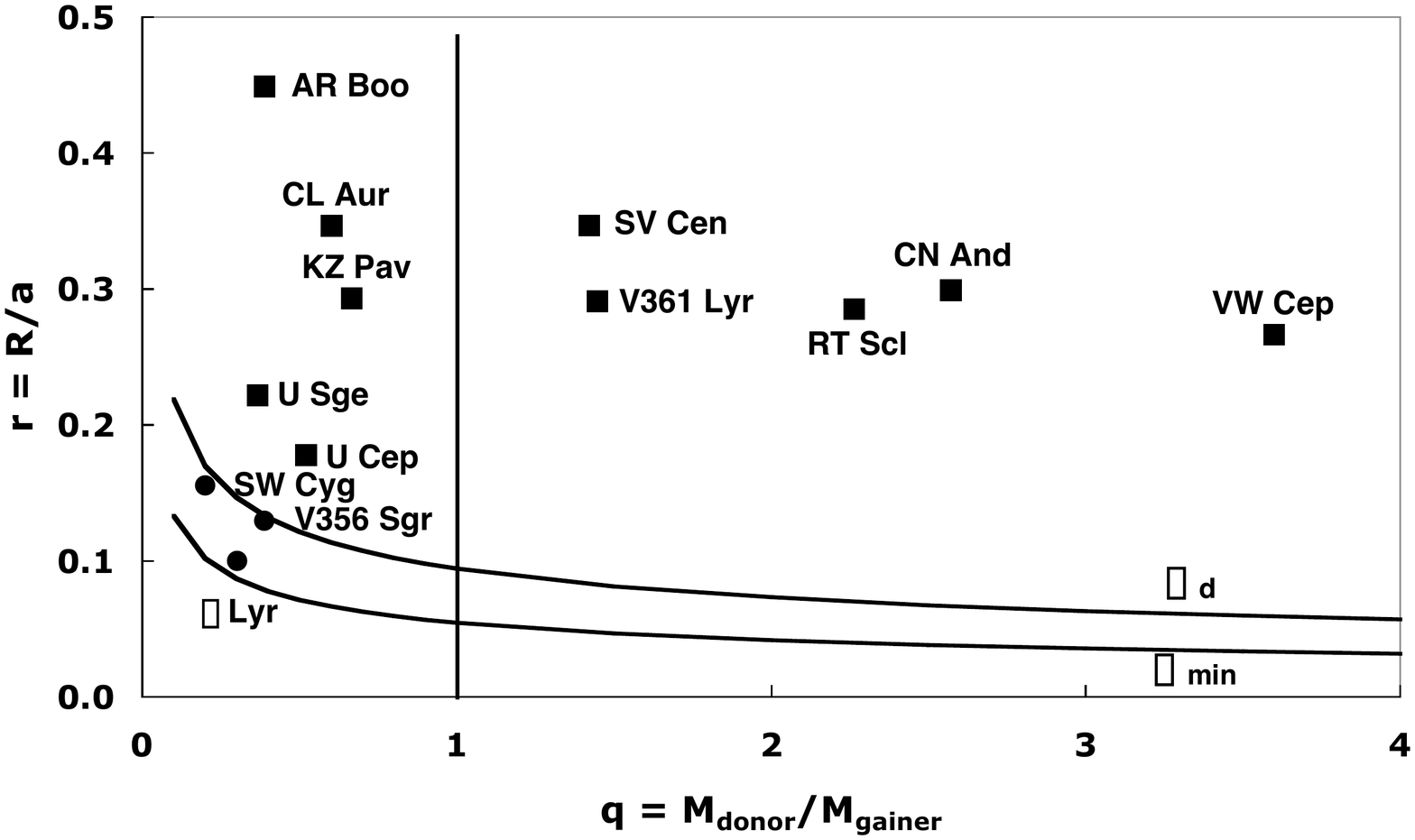}
\caption{($q-r$) diagram for the 13 interacting binaries that were used in our analysis. Algols ($q<1$) are at the left of the vertical line $q$ = 1.0. The ten direct impact systems are located above the curve $\omega_{d}$ (filled squares). The three systems with a transient accretion disk around the gainer are located between the curves $\omega_{min}$ and $\omega_{d}$ (filled circles). Permanent accretion disks would be located below $\omega_{min}$. Systems with $q>1$ have not reached Algol characteristics yet.} 
\label{fig_fig2}
\end{figure*}

There are only 13 reliable hot spot temperatures available in the literature. According to the criterion of Lubow \& Shu (\cite{Lubow}) and as shown in Fig. {\ref{fig_fig2}} ten systems are direct impact systems, while the three others have a transient accretion disk. A permanent accretion disk, emitting continuously in H$\alpha$, can only be formed when the size of the gainer is sufficiently small relative to the size of the orbit.
 
\vspace{0.1cm}

In the case of the formation of a hot spot on the edge of an accretion disk we have to replace in Eq. (\ref{Ktilde}) $R_{g}$ by $R_{disk}$. We further replaced $T_{eff,g}$ by $T_{disk}$ for $\beta~Lyr$ which has an opaque and optically thick accretion disk. The adopted value of $T_{disk}$ $\approx$ 8900 K will be discussed in Sect. \ref{sec_disk}. Two other systems have a transparent and optically thin accretion disk so that their hot spot is seen upon the effective temperature of the gainer star. Table {\ref{tab_tab1}} contains all the data to find a value for $\tilde{K}$ for 13 interacting binaries, with a (rather poor) best fit as a function of the total mass of the system, which is luckily only slightly different from the expression found by Van Rensbergen et al. (\cite{Walter3}) at the time when data were available on only 11 interacting binaries:

\begin{equation}
 \tilde{K} = 3.9188 ~ \left({{\frac{M_{d}}{M_{\odot}}+\frac{M_{g}}{M_{\odot}}}}\right)^{1.645}.
 \label{Ktildenumber} 
\end{equation}

This result allows us to calculate the radiative efficiency $\eta$ of accretion onto a spinning main sequence gainer with a hot spot. Using $L_{add}=L_{acc}^{\infty} ~ D ~  \tilde{K} ~S = \eta~\dot{M}~c^{2}$ with the mass-radius relation determined from the stellar evolutionary models of Schaller et al. (\cite{Schaller}) for gainers in the mass range considered in this paper and in the middle of their main sequence life,  one obtains:

\begin{equation}
 \eta~= 5.85~10^{-6} ~ { \left( \frac{M_{g}}{M_{\odot}  }\right) ^{0.446}} ~ \left({\frac{M_{d}}{M_{\odot}}+\frac{M_{g}}{M_{\odot}}}\right)^{1.645}~D~S.
 \label{etanumber} 
\end{equation}

As expected, we find that this radiative efficiency is low as compared to the values found for accretion onto white dwarfs ($\eta$ $\sim$ $10^{-4}$), neutron stars ($\eta$ $\approx$ 0.2) and black holes ($\eta$ up to 0.423 for the most rapidly rotating black hole).

\begin{table*}
\caption{Data used to determine $\tilde{K}$ for 13 semi-detached binaries.}
\label{table:Ktabel}
\centering
\begin{tabular}{c c c c c c c c c}
\hline
System & $M_{d}+M_{g}$ & $R_{d}+R_{g}$ & $T_{eff,d}$ & $T_{eff,g}$ & $dM/dt$ & $T_{spot}$ & $\tilde{K}$ & $D$\\
 & ($M_{\odot}$) & ($R_{\odot}$) & (K) & (K) & ($M_{\odot}~yr^{-1}$) & (K) & & \\
\hline
VW Cep & 0.90~+~0.25 & 0.93~+~0.50 & 5000 & 5200 & 8.769E-08 & 7076 & 6.528 & 0.0445\\
AR Boo & 0.35~+~0.90 & 0.65~+~1.00 & 5398 & 5100 & 1.484E-07 & 5539 & 1.064 & 0.0535\\
CN And & 1.30~+~0.51 & 1.43~+~0.92 & 6500 & 5911 & 1.215E-07 & 6485 & 8.251 & 0.0261\\
KZ Pav & 0.80~+~1.20 & 1.66~+~1.50 & 5000 & 6500 & 2.629E-08 & 7357 & 12.544 & 0.280\\
V361 Lyr & 1.26~+~0.87 & 1.02~+~0.72 & 6200 & 4500 & 2.178E-07 & 11021& 5.185 & 0.156\\
RT Scl & 1.63~+~0.72 & 1.67~+~1.02 & 7000 & 4800 & 9.500E-08 & 9300 & 36.363 & 0.0923\\
CL Aur & 1.35~+~2.24 & 2.51~+~2.58 & 6323 & 9420 & 1.302E-07 & 10598 & 45.080 & 0.178\\
U Cep & 1.86~+~3.57 & 4.40~+~2.41& 4975 & 11215 & 5.092E-07 & 30000 & 305.631 & 0.577\\
U Sge & 1.99~+~5.45 & 5.64~+~4.11& 5500 & 12250 & 2.036E-06 & 20000 & 49.185 & 0.510\\
SV Cen & 8.56~+~6.05 & 5.90~+~5.00 & 14000 & 23000 & 1.626E-04 & 37580 & 211.238 & 0.0297\\
\hline
System & $M_{d}+M_{g}$ & $R_{d}+R_{disk}$ & $T_{eff,d}$~+~$T_{eff,g}$ & $T_{disk}~+~T_{edge,disk}$ & $dM/dt$ & $T_{spot}$ & $\tilde{K}$ & $D$\\
 & ($M_{\odot}$) & ($R_{\odot}$) & (K) & (K) & ($M_{\odot}~yr^{-1}$) & (K) & & \\
\hline
SW Cyg & 0.50~+~2.50 & 4.30~+~3.44 & 4891~+~9000 & 6308~+~4968 & 2.130E-07 & 13060 & 106.032 & 0.587\\
V356 Sgr & 3.00~+~11.00 & 13.20~+~9.07 & 8600~+~16500 & 6174~+~4299 & 4.442E-07 & 17050 & 659.326 & 0.561\\
$\beta$~Lyr & 4.25~+~14.1 & 16.70~+~15.88 & 13000~+~28000 & 18279~+~8919 & 3.440E-05 & 22590 & 135.355 & 0.446\\
\hline
\end{tabular}
\label{tab_tab1}
\tablefoot{The many digits in the numbers are only significant because they assure the internal consistency between the values of $L_{acc}$, $A$, $S$ and $\tilde{K}$. The first ten binaries are direct impact systems, the last three lines contain disk systems, with radii and temperatures of the gainer and its accretion disk defined separately.}
\end{table*}

\subsubsection{Loss of mass}

The effects of spin-up and accretion luminosity induced photon pressure increase with higher mass accretion rates. The critical mass transfer rate $\dot M_{d}^{RLOF,crit}$ is that at which both effects will just compensate the local gravity, so that the local Eddington luminosity is reached. When the mass loss rate of the donor exceeds this critical value, the gainer will nevertheless accept only a rate $\dot M_{d}^{RLOF,crit}$, and the surplus will be lost from the system, thus resulting in $\beta < 1$. The exact formulation of $\dot M_{d}^{RLOF,crit}$ as a function of evolutionary parameters, as included in the code and thus used to calculate the value of $\beta$ at each moment of the evolution, is derived as equations (21) (for direct impact) and (22) (for systems with an accretion disk) in Van Rensbergen et al. (\cite{Walter2}). These equations imply that each amount of energy extracted from the accretion stream can only be used once, i.e. either for spinning up the gainer or for developing or maintaining its hot spot. It also shows that when $\beta < 1$, this does not necessarily mean that critical rotation is achieved, nor that the accretion luminosity alone exceeds the Eddington luminosity, but that both effects combined exceed the local gravitation.

\subsubsection{Loss of angular momentum}

Our liberal code thus calculates a time dependent $\beta$(t) self-consistently within the model as described in the previous subsection, and assumes that matter is lost from the hot spot on the gainer (or accretion disk around the gainer) so that the escaping matter removes only the angular momentum of the gainer's orbit. However, the near-critical rotational velocity of the gainer considerably enhances its spin angular momentum. The influence of the latter in the angular momentum loss process has not been considered in this study, but will be included in future research. One thus obtains a time dependent value of $\alpha$, describing the angular momentum loss from the system:

\begin{equation}
\alpha = \left(\frac{M_{d}}{M_{g}+M_{d}}\right)^{2}.
\label{alpha} 
\end{equation}
 
A typical value of  $\alpha$ during the fast and liberal era of RLOF A is then $\alpha$~$\approx$~0.25 ($q$~$\approx$~1). When this liberal era is succeeded by a fast and liberal era of RLOF B the value of $\alpha$ turns out to be much smaller and can easily be calculated from relation (\ref{alpha}).
 
\subsubsection{Values and uncertainties for Table {\ref{tab_tab1}}}

Table {\ref{tab_tab1} does not contain the values of the present-day orbital periods of the systems since these values are very well known for the eclipsing binaries used in our analysis. The other values used to construct Table {\ref{tab_tab1}} are discussed in Appendix A.

\vspace{0.1cm}

Relation (\ref{Ktildenumber}) was used in our code to gauge the radiative efficiency of accretion by a main sequence gainer. Figure  \ref{fig_fig3} shows the large uncertainty that remains in the determination of $\tilde{K}$ caused by the errors and uncertainties on the values quoted in Table \ref{tab_tab1}. The curves above and below respectively add and subtract one standard deviation from the values of $\tilde{K}$ as obtained from relation (\ref{Ktildenumber}). It is clear that binary evolution would be more conservative than argued in this paper if the lower curve were valid and less conservative for the higher curve.

\vspace{0.1cm}

Figure \ref{fig_fig3} shows error bars obtained from the data quoted in Appendix A. The calculation of $\tilde{K}$ requires firstly the knowledge of the orbital period and of the masses, radii and effective temperatures of both binary components. The error on the sum of the masses of a binary (horizontal axis of Fig. \ref{fig_fig3}) is not frequently mentioned by the authors. In some other cases an error on the masses could be evaluated from various values found in the literature for a particular system.

\vspace{0.1cm}

Moreover, the calculation of $\tilde{K}$ uses the knowledge of the mass transfer rate ($\dot{M}_{d}^{RLOF}$), the luminosity added to the gainer by the impinging mass ($L_{add}$), the surface of the hot spot ($S_{spot}$) and the hot spot temperature ($T_{spot}$). These observed quantities are theoretically interdependent. The quantity $\tilde{K}$ can e.g. be calculated from Eq. (\ref{Ktilde}) and/or from the relation $\tilde{K}$ = $(A~S)^{-1}$. The quantity $A$ can be determined from observation $A= \frac{L_{acc}} {L_{add}}$ and/or calculated from $S$ and $T_{spot}$. The quantity $S$ can be directly observed and/or calculated from Eq. (\ref{SPringle}). Erroneous evaluations of the observations can therefore be rejected when they violate the required internal consistency, as illustrated in Sect. \ref{sec_Removed}. 
This is the main reason why only 13 binaries could be used in this analysis. Errors on $\tilde{K}$ (vertical axis of Fig. \ref{fig_fig3}) are obtained when the quantities mentioned above can be combined into models that respect every observed quantity within a reasonable range. The length of the vertical error bar then shows the largest and the smallest value that can be calculated for $\tilde{K}$ within this framework. No error is shown in Fig.  \ref{fig_fig3} when the error is too small or when not every quantity has been measured so that the internal consistency could not be completely explored.

\begin{figure*}[!ht]
\centering
\includegraphics[width=9.6cm]{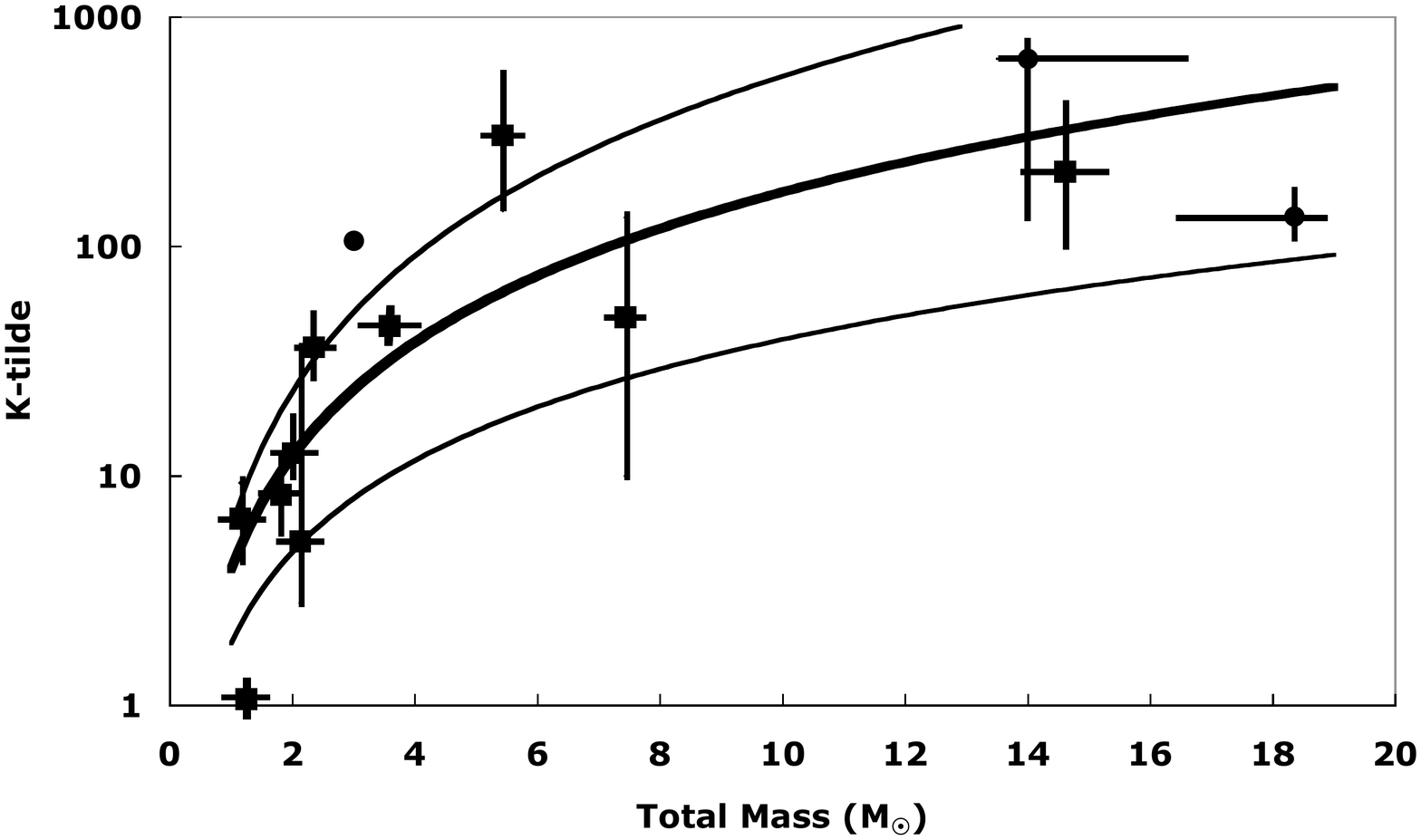}
\caption{Radiative efficiency of accretion $\tilde{K}$ as a function of the total mass of a binary. Relation (\ref{Ktildenumber}) is shown by a thick line. The two thinner lines are at one standard deviation above and below the thick line. Errors and realistic estimates of inaccuracies are shown for individual cases. Filled squares indicate the positions of the ten direct impact systems whereas filled circles are used for the three systems with an accretion disk.}
\label{fig_fig3}
\end{figure*}

\subsection{The distribution of mass ratios and orbital periods of Algols}
\label{sec_section4}

\subsubsection{The observed distributions}
\label{sec_obs}

The distribution of orbital periods of Algols is well established. The distribution of mass ratios is however biased. Van Rensbergen et al. (\cite{Walter2}) used the mass ratio and orbital period distribution of Algols out of a sample of 303 observed systems. These systems were taken from the catalog of Budding et al. (\cite{Budding et al.}) extended with semi-detached Algols from the catalog of Brancewicz et al.  (\cite{Brancewicz et al.}). All these systems  are issued from a binary with a B-type primary at birth. A majority of 268 systems have a late B-type primary progenitor, leaving only 35 systems with an early B-type primary at birth. When discussing initial conditions (e.g. in Sections \ref{sec_Initial} and \ref{sec_Amount}), we use $q$ = ${M_{g}\over M_{d}}$ as definition of the mass ratio. However, since in Algols the gainer has become the most massive component, \emph{we use $q$ = ${M_{d}\over M_{g}}$ as the definition of the mass ratio of an Algol system}. 

\subsubsection{The calculated distributions}
\label{sec_calc}

Using the initial conditions of Sect. \ref{sec_Initial} and the data of our library with liberal binary evolutionary calculations yielding the durations that an Algol lives with well defined values of the orbital period and mass ratio, Van Rensbergen et al. (\cite{Walter3}) used a Monte Carlo simulation to calculate distributions of mass ratios and orbital periods of Algols that started RLOF during core hydrogen burning of the donor (case A). This paper extends these calculations with Algols issuing from binaries that started RLOF during hydrogen shell burning of the donor (case B). We compare the results obtained under the assumption of strong tidal interaction with those obtained with weak tides. Although significant differences frequently occur between both assumptions, they yield a very similar overall distribution of mass ratios and orbital periods of Algols.

\vspace{0.1cm}

Systems which become too faint with increasing distance to be included in the catalogs are removed by our Monte Carlo simulation. Therefore we added the spatial distribution of binaries into the simulation so that they have either a spherically uniform distribution around the Sun or that they are uniformly distributed in a flat structure such as Gould's belt. These two different assumptions yield also however no significant nor systematic differences.

\section{Results}

\subsection{Calculated mass driven out of a close binary}
\label{sec_Amount}

Using the stellar evolution code to calculate the evolution of binaries including those performing case B RLOF, we find that the amount of mass driven out of a close binary ($\Delta$$M$) increases with increasing initial mass of the donor and with increasing initial orbital period. Binaries having an initial mass of the donor star below 6 $M_{\odot}$ lose no or almost no matter during their evolution. This is because the mass transfer rate in these systems never exceeds the critical value which is required in order for the combination of spin-up and accretion luminosity to overcome the gravitational potential of the system. Binaries with donors having initial masses between 6 $M_{\odot}$ and 9 $M_{\odot}$ undergo a moderate amount of mass loss starting at larger initial orbital periods. Binaries with an initial mass of the donor star above 9 $M_{\odot}$ suffer from severe mass loss during their evolution within this scenario. Although large individual differences may occur between binaries evolving with strong, respectively weak, tidal interaction, there are no significant differences between the two groups. Figure \ref{fig_fig4} shows an example of the amount of mass lost by a binary with a 7 $M_{\odot}$ primary at birth and evolving with strong tidal interaction. The grid of calculated cases covers initial mass ratios $q$ = 0.4, 0.6 and 0.9.  At small initial orbital periods no mass is lost because RLOF is ignited before the donor has become large enough to transfer mass to the gainer at a rate larger than the critical one. At higher initial orbital periods most mass is lost in the $q$ = 0.6 case, sufficiently large to ensure a large enough separation (according to Kepler's third law) so that the mass transfer rate is higher than with $q$ = 0.4. The mass lost in the case of $q$ = 0.9 is however smaller because such systems grow into near contact more easily, leading to a small geometrical factor $D$ and thus to a fading accretion luminosity. The largest amounts of mass are lost in those cases where RLOF starts after exhaustion of hydrogen in the core of the donor star. This will be achieved in this case with initial orbital periods above $\sim$ 4 days.  The liberal calculations in our catalog (Van Rensbergen et al. \cite{Walter2}) give also the amount of mass that is lost by many binaries with initial mass of the primary between 3 $M_{\odot}$ and 15 $M_{\odot}$.

\begin{figure*}[!ht]
\centering
\includegraphics[width=9.6cm]{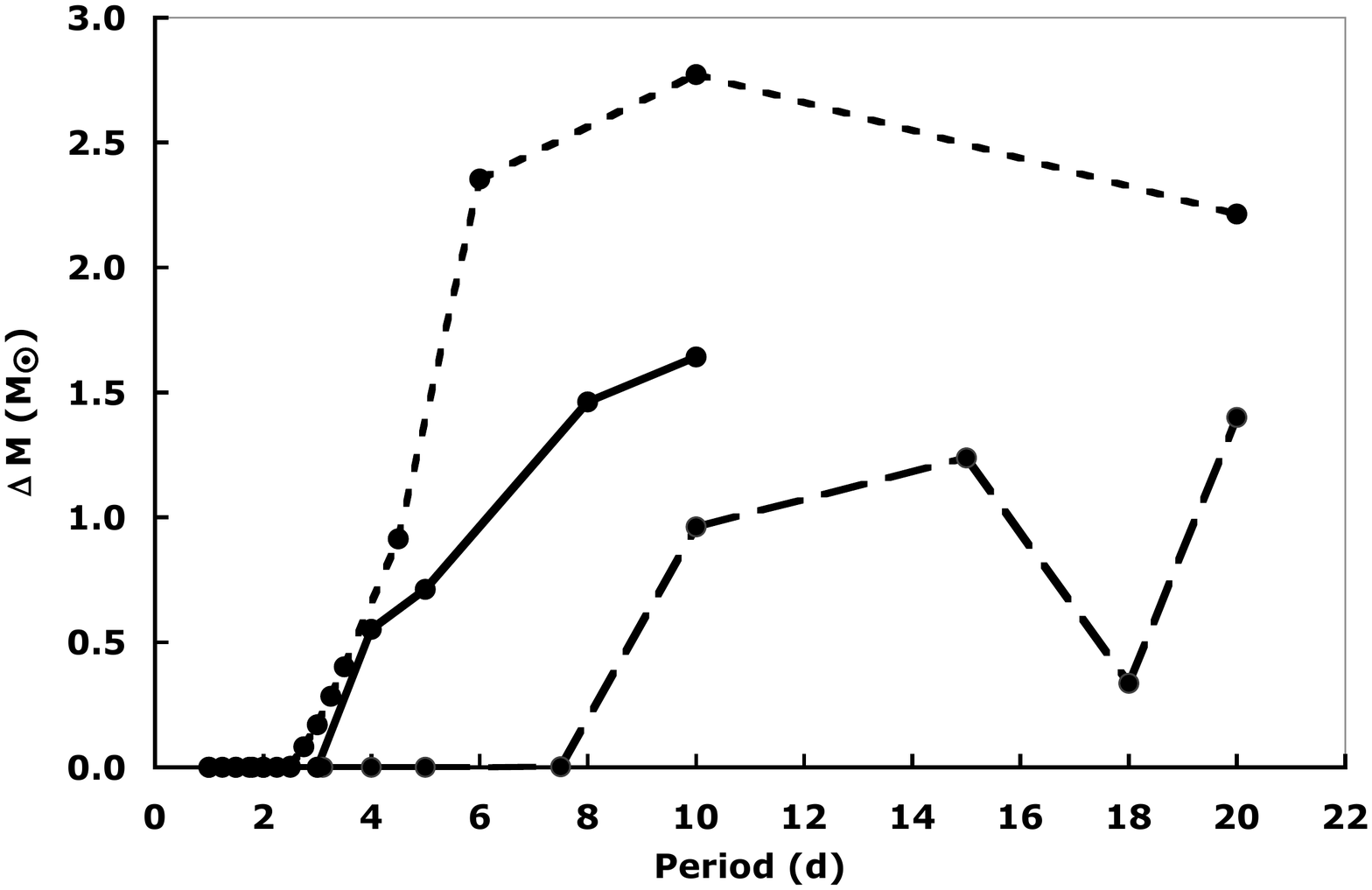}
\caption{Calculated amounts of mass lost by a binary with a 7$M_{\odot}$ primary at birth undergoing strong tidal interaction during its evolution. The total amount of lost mass $\Delta$$M$ is on the vertical axis, the initial orbital period is on the horizontal axis. Calculated cases are indicated by filled circles. Binaries starting with an initial mass ratio $q$ = 0.4 are connected by a long-dashed line, those with $q$ = 0.6 by a dashed line and those with $q$ = 0.9 by a solid line.} 
\label{fig_fig4}
\end{figure*}

\subsection{Period distribution of Algols with a B-type primary at birth}
\label{sec_period}

Figure {\ref{fig_fig5}} compares the observed orbital period distribution of Algols with a B-type primary at birth with theoretical simulations as obtained with the liberal assumptions. This distribution is not affected by the different assumptions that were made about the tidal interaction. Algols observed with orbital periods of more than 15 days which were not reproduced sufficiently by considering the case A RLOF alone (Van Rensbergen et al. \cite{Walter3}) are very well represented now that case B is included in the calculations. The maximum deviation in a one-sample Kolmogorov-Smirnov test decreases moreover from 0.115 obtained with the conservative simulation to the significantly better value of 0.067 when the liberal simulation is used. The observed period distribution of Algols is however only very closely reproduced by the liberal simulation for periods above 1 day. The number of $\sim$ 9 $\%$ of Algols observed with orbital periods below 1 day is largely underestimated by our simulation because these binaries tend to merge in our calculations. The fact that their number is very well observed could mean that binaries can survive overcontact better than tolerated by our code (see discussion in Sect. \ref{sec_Discussion}).

\begin{figure*}[!ht]
\centering
\includegraphics[width=9.6cm]{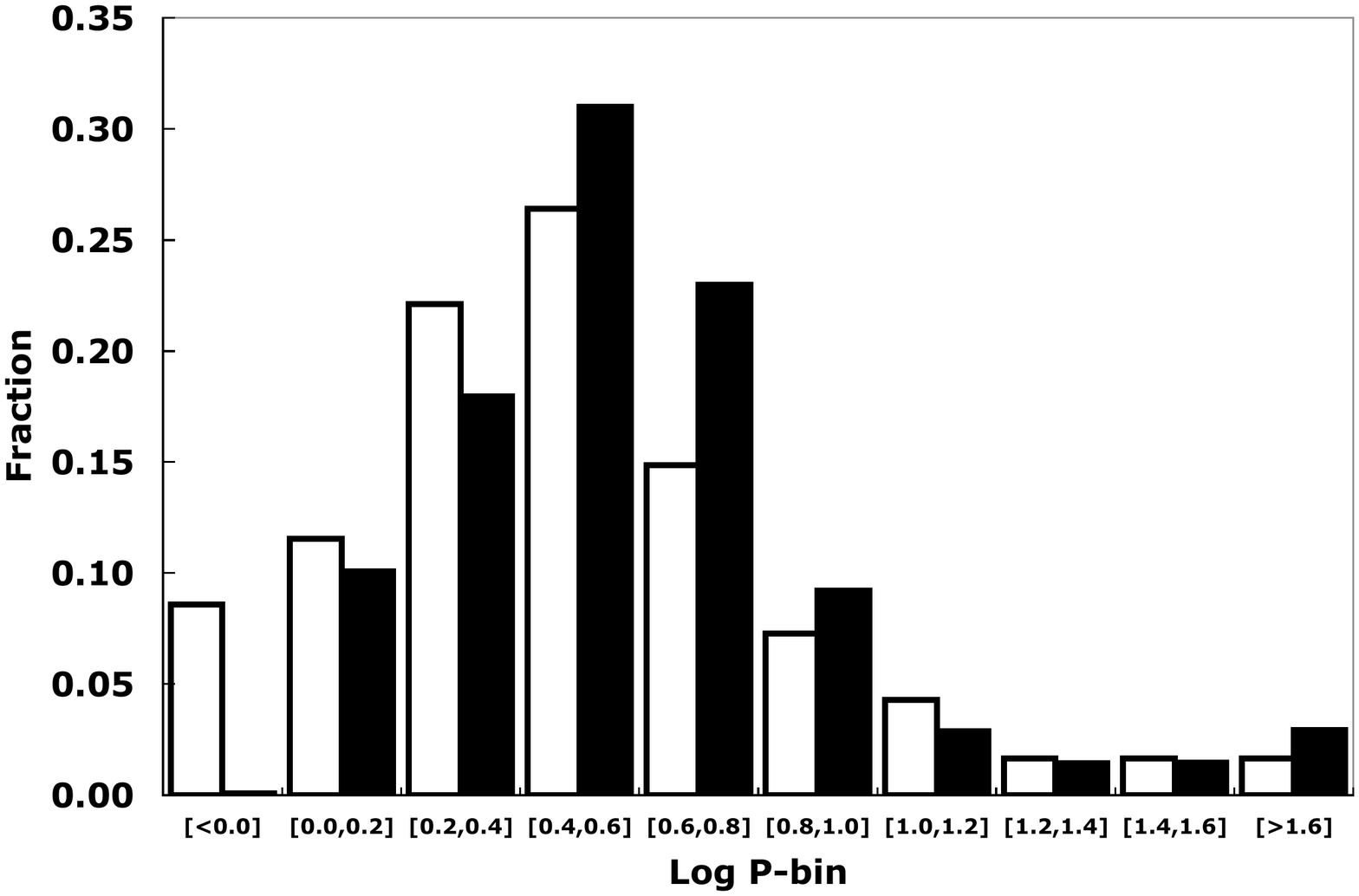}
\caption{Observed distribution (white) of orbital periods of 303 Algols as compared to liberal (black) simulation.} 
\label{fig_fig5}
\end{figure*}

\subsection{The mass ratio distribution of Algols with a B-type primary at birth}
\label{sec_late}

We used an observed distribution of Algols with a B type primary at birth, having $\sim$ 45 $\%$ in the high mass ratios ($q$ $\in$ [0.4,1]). More specifically we find $\sim$ 24 $\%$ with $q$ $\in$ [0.4,0.55] and $\sim$ 21 $\%$ with $q$ $\in$ [0.55,1]. The fact that the conservative binary evolutionary  code did not reproduce this observed mass ratio distribution for a sample of 303 Algols was the initial reason for us to start thinking about a liberal scenario. 

Van Rensbergen et al. (\cite{Walter1}) used $\beta$ as a parameter taken equal to 0.5 during the entire binary evolution. They found that this a priori  $liberal$ assumption fits the observed mass ratio distribution better than the $conservative$ assumption does. In our subsequent papers we upgraded the quantity from a free constant parameter to a time dependent quantity $\beta$(t) which is determined by the physical characteristics of the system.

\vspace{0.1cm}

Conservative codes produce only $\sim$ 12 $\%$ of Algols with a mass ratio $q$ above 0.4.  Including case B binary evolution to our statistics, this fraction rises up to  $\sim$ 17 $\%$. Both results are shown in Fig. {\ref{fig_fig6}}. It is clear that this limited enhancement is entirely due to those binaries with an initial primary mass above  $\sim$ 7 $M_{\odot}$ because in our scenario the less massive systems hardly lose any mass during their evolution. Moreover only $\sim$ 2 $\%$ of the Algols are reproduced with a mass ratio above 0.55, a number which is largely inferior to the $\sim$ 21 $\%$ required by the observations.

\begin{figure*}[!ht]
\centering
\includegraphics[width=9.6cm]{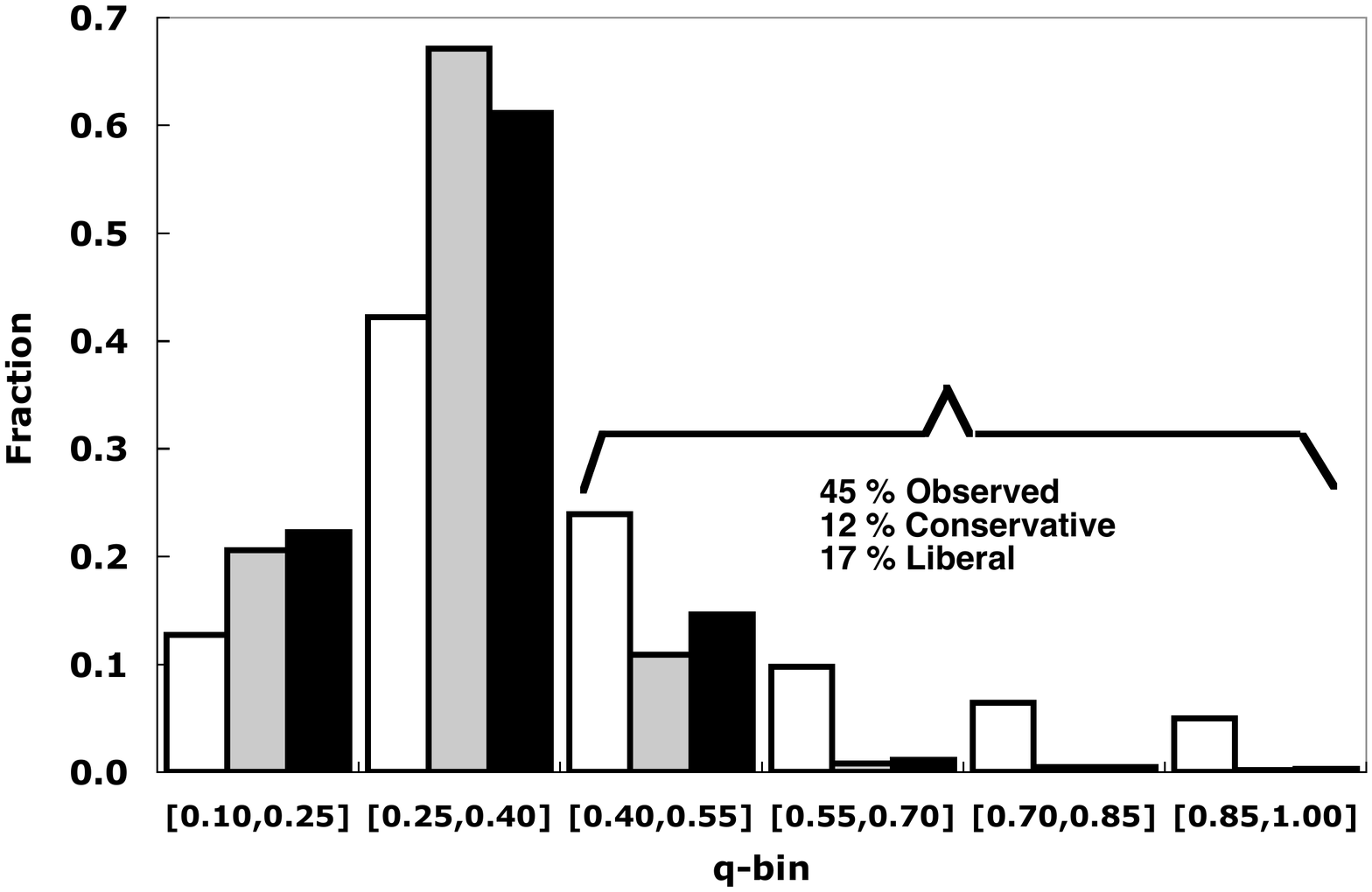}
\caption{Observed distribution of mass ratios of 303 Algols (white) as compared to conservative (grey) and liberal (black) simulation.} 
\label{fig_fig6}
\end{figure*}

In Sect. \ref{sec_Amount} we showed that binaries with an early B-type primary at birth lose a large fraction of the transferred mass during their evolution. The short liberal era during which mass is lost ($\beta < 1$) occurs when the mass transfer rate is large, i.e. soon after the onset of RLOF when the binary is in its pre- or early-Algol stage. The mass ratio distribution as obtained with the liberal assumptions for binaries with an early B-type primary at birth will differ very much from the one obtained using the conservative binary evolutionary code. Hence we obtain $\sim$ 33 $\%$ of Algols with $q$ $\in$ [0.4,0.55] when we consider the liberal evolution of binaries with a mass of the initial primary above 7 $M_{\odot}$. Above 8 $M_{\odot}$ this fraction even rises up to $\sim$ 45 $\%$. The fraction with $q$ $\in$ [0.55,1] remains however always in the vicinity of 2 $\%$, a number which is far below what is required by our set of observations.

\section{Discussion}
\label{sec_Discussion}

\begin{itemize}

\item Since more Algols are observed with orbital periods below one day than are produced by either evolutionary code, we propose that theoretical evolutionary calculations for binaries should allow for the survival of the system even after deep contact. This is similar to stating that such systems seem to survive a specific kind of common envelope evolution involving two main sequence stars. The nature of such evolution requires further exploration, as the physics are probably different from and more complex than the physics in the classical common envelope scheme as described by e.g. Webbink (\cite{Webbink}) and Nelemans \& Tout (\cite{Nelemans}).

\item Although the theoretically calculated liberal mass ratio distribution of Algols with an initial early B-type primary fits the observations better, the overall observed mass ratio distribution of Algols still shows too many systems with high mass ratios. This could partly be due to theoretical mass transfer rates which are too large and too short living during the rapid era of mass transfer. We have compared mass transfer rates as obtained by our binary evolutionary code in the conservative mode with mass transfer rates for conservative evolution as produced by previous authors and published by Kippenhahn $\&$ Weigert (\cite{Kippenhahn1}), Kippenhahn et al. (\cite{Kippenhahn2}) and Paczy\'nski (\cite{Paczynski1}, \cite{Paczynski2}). The durations of our calculated phases of rapid mass transfer are very similar to those mentioned above, whereas our peak values are somewhat lower because our stellar models are calculated with OPAL opacities (Rogers \& Iglesias \cite{Rogers}, Iglesias et al. \cite{Iglesias1}, Iglesias \& Rogers \cite{Iglesias2}), which were not available previously. Our calculated durations and peak values of mass transfer rates agree very well with those published by Nelson $\&$ Eggleton (\cite{Nelson}) for a representative set of interacting binaries. The occurrence of many observed Algols with high mass ratios thus remains unexplained.
\item Our scenario predicts only $\sim$ 2 $\%$ of Algols with a mass ratio above 0.55. It is not certain that the discrepancy with $\sim$ 21 $\%$ of observed Algols with these large mass ratios is entirely due to fundamental failures of the theory. We would suggest that the determination of mass ratios from observations is open for revision. Algols with a high mass ratio are present mainly when determined using main sequence characteristics for the gainer ($q_{MS}$). Other methods generate an observed mass ratio distribution with much less high values (see Sect. 3 of Van Rensbergen et al. \cite{Walter1}). The validity of one method over another should be clearly decided before any conclusions can be attached to the difference between observed and predicted number of binaries in these high-q-bins.
\item Call for observations: Much work has to be done in the future on the calibration of the efficiency by which mechanical energy is transformed into radiation as represented by relation (\ref{Ktildenumber}). This can only be done when data for Table \ref{tab_tab1} can be refined and extended. The crucial quantity $\tilde{K}$ is on one side determined by relation (\ref{Lacc}) and on the other side by relation (\ref{Ktilde}).

\vspace{0.1cm}

Relation (\ref{Lacc}) uses the product of $A >$ 1 with $S \ll$ 1. The quantity $A$ is determined from the knowledge of $L_{acc}$ and $L_{add}$. Assuming that the geometry of the binary is perfectly known, the value of $L_{acc}$ is determined with the (O-C) method (Sterken \cite{Chris}) yielding the mass transfer rate only when the ephemeris of the eclipse changes linearly with time. The luminosity added to the gainer $L_{add}$ has been determined in only a few cases. The fractional surface has only been measured sometimes since small spot radii ($< 10^{\circ}$) are hard to observe (Van Hamme et al. \cite{Vanhamme}). Fortunately, we have at our disposal the theoretical relation as derived by Pringle (Eq. (\ref{SPringle})) which has no observational limitations and which is reliable within a factor of two for every case that could be checked in this paper.

\vspace{0.1cm}

Relation (\ref{Ktilde}) uses quantities which are relatively well known such as the radius of the gainer and its effective temperature. The uncertainty on the quantity $L_{acc}$ is discussed above. Measurements of the spot temperature $T_{spot}$ are crucial and are given in Sections  \ref{sec_direct} and  \ref{sec_disk}.

Binaries for which the value of $\tilde{K}$ as determined from relation (\ref{Lacc}) is not compatible within resonable limits on the uncertainties on $S$, $L_{add}$, $L_{acc}$, $T_{spot}$ with the value as determined from relation  (\ref{Ktilde}) have been removed from our sample.

Although masses, radii and effective temperature of donor and gainer are usually well known, there is an explicit need for more and more reliable determinations of characteristics of the gainer star in semi-detached binaries. Observable quantities such as $S$, $L_{add}$, $L_{acc}$, $T_{spot}$ are needed to construct a better model which allows some binaries to lose mass into interstellar space.

\end{itemize}

\section{Conclusions}

\begin{itemize}

\item A catalog (Van Rensbergen et al. \cite{Walter2}) with 561 conservative and liberal evolutionary tracks is available at the Centre de Donn\'ees Stellaires (CDS). Binaries with an initial primary mass $\in$~[3,5]~$M_{\odot}$ are calculated in one mode only since they evolve conservatively. Binaries with an initial primary mass $\in$~[6,15]~$M_{\odot}$ are calculated in the liberal mode. Results for evolution with weak and strong tidal interaction are given separately. Conservative tracks are always added so that the reader is able to compare results of liberal (in two different tidal modes) and conservative evolution.
\item Mass impinging from the donor spins up the gainer and creates a hot spot in its equatorial zone or at the edge of its accretion disk. The combined energy of the enhanced rotation and increased radiation from the hot spot may exceed the binding energy of the system. We present the results of liberal evolutionary calculations for binaries with a B-type primary at birth. A significant fraction of the transferred mass is lost by the system in the case of binaries with an early B-type primary at birth. Systems with a late B-type primary at birth hardly lose any matter. 
\item The observed distribution of orbital periods of Algols is well reproduced by our liberal theoretical calculations. Algols with an orbital period smaller than 1 day are however present for $\sim$ 9 $\%$ in the observations whereas they are almost not created by our simulation. Our liberal theoretical calculations reproduce the adopted observed distribution of mass ratios of Algol binaries less satisfactorily: Algols with mass ratios larger than 0.55 are underrepresented. Sect. \ref{sec_Discussion} proposes possible additions to the present scenario to meet these discrepancies.

\end{itemize}

\begin{acknowledgements}

We thank Ed van den Heuvel for his permanent interest in our work and his fruitful comments, Walter van Hamme for his clarifying comments on the measurement of hot spot sizes and Chris Sterken to make us acquainted with the (O-C) method which enabled us to calculate and confirm many mass transfer rates.

\end{acknowledgements}

    \begin{appendix}
    \label{sec_appendix}
   
   \section{Data for 13 semi-detached binaries}
   
   \subsection{Ten systems with direct impact}
   \label{sec_direct}

Data on present masses and radii of the stars, mass transfer rates, temperature of the hot spot ($T_{spot}$), fractional size of the hot spot ($S$), luminosity added by the hot spot ($L_{add})$ were taken for the ten systems with direct impact on the gainer's surface from the following references:

\begin{itemize}
\item VW Cep: Masses, radii, effective temperatures of donor and gainer are from Pustylnik  $\&$ Niarchos (\cite{Pustylnik}). $L_{add}$, $T_{spot}$, $R_{spot}$ (and thus the quantity $S$) on the gainer are from the same authors. 
Values for ${\dot P}$ and thus of the mass transfer rate ${\dot M}$ are from Pribulla et al. (\cite{Pribulla}) and Devita et al. (\cite{Devita}).
\item AR Boo: Masses, radii, effective temperatures of donor and gainer and also values for $T_{spot}$ and ${\dot M}$ are from Lee et al. (\cite{Lee1}).
\item CN And: Masses, radii, effective temperatures of donor and gainer and also values for $T_{spot}$,  $S_{spot}$ and ${\dot M}$ are from Van Hamme et al. (\cite{Vanhamme}) and references therein. A mean value of ${\dot M}$ was taken by considering also a somewhat different value of ${\dot M}$ determined by Samec et al. (\cite{Samec}) and used by Qian (\cite{Qian}) in his analysis of near-contact binary systems.
\item KZ Pav:  Masses, radii, effective temperatures of donor and gainer and also values for $L_{add}$ and ${\dot P}$ are from Budding et al. (\cite{Budding}). We used the latter value of  ${\dot P}$ in order to calculate  ${\dot M}$.
\item V361 Lyr: Masses, radii, effective temperatures of donor and gainer are given by Hilditch (\cite{Hilditch}) and Yakut $\&$ Eggleton (\cite{Yakut}). Hilditch (\cite{Hilditch}) also gives values for $T_{spot}$, $L_{add}$ and ${\dot M}$. This binary is thus very completely described.
\item RT Scl:  Masses, radii, effective temperatures of donor and gainer are given by Banks et al. (\cite{Banks}) and Pourbaix et al. (\cite{Pourbaix}). Banks et al. (\cite{Banks}) also give a value for $T_{spot}$. Values for ${\dot M}$ are given by Clausen $\&$ Gronbech (\cite{Clausen}) and Duerbeck $\&$ Karimie (\cite{Duerbeck}). From this we calculated a mean value for ${\dot M}$ which is a factor of $\sim$ 2 smaller than the value given by Rafert $\&$ Wilson (\cite{Rafert}).
\item CL Aur:  All the data for this Algol binary, including a value for $T_{spot}$ on the gainer are given by Lee et al. (\cite{Lee2}).
\item U Cep:   Masses, radii, effective temperatures of donor and gainer are given by Budding et al. (\cite{Budding et al.}). A value for  ${\dot P}$=+9.796~$10^{-7}$$d$$y^{-1}$ was derived by ourselves from the ``Eclipsing Binaries Minima Base'' of Kundera, completed with the epochs of the most recent eclipses that can easily be found on the CDS website. From this we derived a value of ${\dot M}$ which is a factor of $\sim$ 2 smaller than the evaluation by Pustylnik (\cite{Pustylnik2}). The value of $T_{spot}\in [20000,40000]$~K was determined by Peters (\cite{Peters2}) from the observations with the FUSE satellite. A refined value of $T_{spot} \approx 30000$~K was then adjudged to the hot spot by the same author if one assumes that this HTAR (High Temperature Accretion Region) radiates as a black body. It is also interesting to notice that the gainer in this binary rotates 7.8 times faster than the synodic value (Dervisoglu et al. \cite{Dervisoglu}).
\item U Sge:  Masses, radii, effective temperatures of donor and gainer are given by Kempner $\&$ Richards (\cite{Kemner}) and Vesper et al. (\cite{Vesper}). Manzoori $\&$ Gozaliazl (\cite{Manzoori}) have determined a value for ${\dot P}$ and thus ${\dot M}$ and have seen a hot spot on the gainer's surface for which a range of temperatures has been given by Richards $\&$ Albright (\cite{Richards}). A value for $L_{add}$ has been determined by Albright $\&$ Richards (\cite{Albright}). This makes this binary also a well described system.
\item SV Cen: Masses and radii of donor and gainer are given by Rucinski et al. (\cite{Rucinski}) because this author considers SV Cen not as a contact binary but as a semi-detached binary. The effective temperatures of both components are taken from the calibration for a B6.5III-donor and a B2V-gainer. Although Wilson $\&$ Starr (\cite{Wilson}), Drechsel et al. (\cite{Drechsel}) and Herczeg $\&$ Drechsel (\cite{Herczeg}) consider SV Cen as an over-contact binary, many of their observations have been interpreted with SV Cen as a very active interactive binary. Herczeg $\&$ Drechsel (\cite{Herczeg}) report a large but variable value of ${\dot P}$. Large mass transfer rates of $\sim$ $4~ 10^{-4} M_{\odot}~yr^{-1}$ and ~``a~few~times''~$10^{-4} M_{\odot}~yr^{-1}$ have been calculated by Wilson $\&$ Starr (\cite{Wilson}) and Rucinski et al. (\cite{Rucinski}) respectively. Moreover a very large value for the hot spot temperature ($T_{spot} \approx 10^{5}$~K) and a very small value for the relative size of the hot spot ($S<$ 0.01) have been reported by Drechsel et al. (\cite{Drechsel}). In order to keep the values of ${\dot M}$ (and thus $L_{acc}$), $A$, $S$ and $\tilde{K}$ internally consistent and physically acceptable, we used in Table {\ref{tab_tab1}} the value of ${\dot M}~=~1.608~ 10^{-4} M_{\odot}~yr^{-1}$ as given by Hilditch (\cite{Hilditch}), leading to a spot temperature of $\sim$ 37500~K which is well below the extremely large value of $\sim$ $10^{5}$~K cited above.
\end{itemize}

\subsection{Three gainers with an accretion disk}
\label{sec_disk}

We have three disk systems in our sample. A hot spot temperature is available for $\beta$ Lyr only. The quantity $L_{add}$ which is needed to calculate the hot spot temperature is calculated with:

\begin{equation}
\left (\frac{L_{add}}{L_{\odot}}\right) = \left (\frac{S_{spot}}{S_{\odot}}\right)~ \left (\frac{T_{spot}^{4}-T^{4}}{5770^{4}}\right).
\label{Ladd} 
\end{equation}

The hot spot is now on the edge of the disk and its surface has never been measured. Expression (\ref{SPringle}) was derived for the calculation of the size of a hot spot on the surface of the gainer and can thus hardly be used in this case. Using this expression nevertheless in the case of $\beta$ Lyr we find an amount $L_{acc}$ that exceeds the calculated accretion luminosity by a factor of three. In order to obtain the added luminosity we have therefore divided the surface area of the impact zone as given by Pringle (\cite{Pringle}) by the same factor of three for the three cases in this subsection. In equation (\ref{Ladd}) we used the effective temperature of the gainer for the two systems where the hot spot is located nearby the surface of the gainer (SW Cyg and V356 Sgr) and we used the temperature in the middle of the accretion disk where the spot is visible as a source of bipolar jets for $\beta$ Lyr (Linnell \cite{Linnell}). In order to calculate $\tilde{K}$ we replaced in relation  (\ref{Ktilde}) $R_{g}$ by $R_{disk}$ and $T_{eff,g}$ by $T_{edge,disk}$, using relations (\ref{Disktemp}) and (\ref{Disktemprad}).

\vspace{0.1cm}

The characteristic disk temperature and the temperature distribution from the stellar surface to the edge of the disk were calculated from the model of Bath \& Pringle (\cite{Bath}). Horne (\cite{Horne}) developed a method of eclipse mapping of accretion disks with this model and Rutten et al. (\cite{Rutten}) applied the latter method to construct models of accretion disks for six cataclysmic variables. The values of $T_{disk}$ and $T_{disk,edge}$ mentioned in Table \ref{tab_tab1} were thus calculated with: 

\begin{equation}
T_{disk}~=\left (\frac{3~G~M_{g}\left \vert {\dot M}\right \vert}{8~\pi~\sigma_{R}~R_{g}^{3}}\right)^{\frac{1}{4}}~=~478074~\mathrm{K}~\left (\frac{M_{g}\left \vert {\dot M}\right \vert}{R_{g}^{3}}\right)^{\frac{1}{4}}
\label{Disktemp} 
\end{equation}

with $\sigma_R = 5.67~10^{-5} erg~cm^{-2}K^{-4}s^{-1}$ the Stefan-Boltz\-mann constant and

\begin{equation}
T(r)~=~T_{disk}~\left (\frac{R_{g}}{r}\right)^{\frac{3}{4}}.
\label{Disktemprad} 
\end{equation}

For the three systems with an accretion disk around the gainer, data on present masses and radii of the stars, mass transfer rates, temperatures of the hot spot ($T_{spot}$) and added luminosities through the hot spot ($L_{add}$) were taken from the following references:

\begin{itemize}
\item SW Cyg: Masses, radii and effective temperatures can be found in the catalog of Budding (\cite{Budding et al.}) and Richards $\&$ Albright (\cite{Richards}). Values for ${\dot P}$ (and thus ${\dot M}$) are from Qian et al. (\cite{Qian2}). Albright $\&$ Richards (\cite{Albright}) argue that the luminosity added to the similar system TT Hya by accretion is five times as large as compared to SW Cyg. Since Peters $\&$ Polidan (\cite{Peters4}) establish a hot spot temperature of 17000~K in TT Hya (more than 60 $\%$ hotter than the underlying effective temperature of the gainer), we conclude that the obtained hot spot temperature of $\sim$ 13000~K for SW Cyg is reasonable.
\item V356 Sgr: Masses and radii of donor and gainer were found by Peters $\&$ Polidan (\cite{Peters3}). These authors also mention the existence of very hot circumstellar material around this binary and of a hot spot concentrated near the photosphere of the B-type gainer. Effective temperatures of donor and gainer are from Polidan (\cite{Polidan}). We derived a value of ${\dot M}$ from the quadratic term in the ephemeris of eclipses as given by Hall et al. (\cite{Hall}). This value is in good agreement with the mass transfer rate as given by Polidan (\cite{Polidan}). It is pointed out by Simon (\cite{Simon}) and Dervisoglu et al. (\cite{Dervisoglu}) that the gainer rotates at the the large equatorial velocity of  $\sim$ 212 $km~s^{-1}$. This is however far below the critical velocity of $\sim$ 612 $km~s^{-1}$ so that in view of the Roche model (e.g. Maeder \cite{Maeder}) the gainer remains practically spherical, and the observed UV-excess cannot be due to the oblateness of the gainer with polar regions much brighter than the equatorial ones, as suggested by Polidan (\cite{Polidan}).
\item $\beta$ Lyr: Masses and radii of donor and gainer are from Simon (\cite{Simon}). Effective temperatures are from Harmanec (\cite{Harmanec}). We derived a mass transfer rate from the values of ${\dot P}$ given by Simon (\cite{Simon}) and Ak and al. (\cite{Ak}). Harmanec (\cite{Harmanec}) reveals a minimum temperature for a hot spot of 20000~K. Linnell (\cite{Linnell}) proposes that the kinetic energy of the mass transfer stream is converted into heat and radiation at a radius equal to half of the radius of the accretion disk. Schmidtt et al. (\cite{Schmidtt}) observed strong H$\alpha$ originating from the disk, possibly at the point of impact of the gas in the disk. The disk temperature of 8500 K as mentioned by  Harmanec (\cite{Harmanec}) is very similar to the temperature we find at the edge of the opaque accretion disk of $\beta$ Lyr ($\sim$ 8920~K). The masses of the components as derived by Zhao et al. (\cite{Zhao}) are somewhat lower than those mentioned by  Simon (\cite{Simon}). They yield somewhat lower values of the disk temperatures and somewhat larger values for $\tilde{K}$. They are included in the error bars in Fig. {\ref{fig_fig3}}.
\end{itemize}

\subsection{Systems removed from the sample}
\label{sec_Removed}

We had to exclude many binaries from our statistical analysis. Some systems are indeed extensively studied with one crucial quantity that was however never determined from the observations. For other systems we found observational values of physical quantities which were contradictory. We mention a few examples of systems which were for the time being excluded from our statistics.

\begin{itemize}
\item TT Hya: Many data on this Algol system with an accretion disk are known, including the value of 17000~K for the hot spot temperature by Peters $\&$ Polidan (\cite{Peters4}). The mass transfer rate of $\geq$~2~$10^{-10} M_{\odot}~yr^{-1}$ as found by Miller et al. (\cite{Miller}) is however five orders of magnitude too low in order to obtain such a high hot spot temperature. The quoted value of the mass transfer rate would yield a hot spot with a temperature of 1000~K at most: only a little bit higher than the temperature at the edge of the accretion disk.
\item $\beta$ Per: The same can be said about the one and only real Algol, $\beta$ Per. Images of accretion structures containing a stable emission source on the gainer were obtained with Doppler tomography by Richards (\cite{Richards1}). A high temperature region (T $\approx$ $10^{4}$ - $10^{5}$~K) surrounding the gainer has been derived from the observation of high temperature resonance emission lines by Cugier \& Molaro (\cite{Cugier}) and Brandi et al. (\cite{Brandi}). Blondin et al. (\cite{Blondin}) found that the gas stream towards the gainer is heated up to a temperature of 2 to 4~$10^{6}$~K for this direct impact system. Such high temperatures can not be produced by a mass transfer rate as low as 0.4 to 2 $~10^{-11} M_{\odot}~yr^{-1}$ as proposed by Pustylnik (\cite{Pustylnik2}). A hot spot with a temperature of $\sim$~$10^{5}$~K can only be created through a mass transfer rate which is seven orders of magnitude larger. A mass transfer rate as low as $\sim$~$10^{-11} M_{\odot}~yr^{-1}$ would raise the photospheric temperature of the gainer with less than 1~K according to the scenario used to construct Table \ref{tab_tab1}.
\end{itemize}

 \end{appendix}
   
\end{document}